Abu Hanif Muhammad Syarubany, 유창동

한국과학기술원

e-mail：hanif.syarubany@kaist.ac.kr

# PentaRAG: Large-Scale Intelligent Knowledge Retrieval for Enterprise LLM Applications

Abu Hanif Muhammad Syarubany and Chang Dong Yoo

Korea Advanced Institute of Science & Technology

## Abstract

Enterprise deployments of large-language model (LLM) demands continuously changing document collections with sub-second latency and predictable GPU cost requirements that classical Retrieval-Augmented Generation (RAG) pipelines only partially satisfy. We present PentaRAG, a five-layer module that routes each query through two instant caches (fixed key–value and semantic), a memory-recall mode that exploits the LLM's own weights, an adaptive session memory, and a conventional retrieval-augmentation layer. Implemented with Ministral-8 B, Milvus and vLLM, the system can answer most repeated or semantically similar questions from low-latency caches while retaining full retrieval for novel queries. On the TriviaQA domain, LoRA fine-tuning combined with the memory-recall layer raises answer similarity by $\approx 8\%$ and factual correctness by $\approx 16\%$ over the base model. Under a nine-session runtime simulation, cache warming reduces mean latency from several seconds to well below one second and shifts traffic toward the fast paths. Resource-efficiency tests show that PentaRAG cuts average GPU time to $0.248\,\text{s}$ per query, roughly half that of a naïve RAG baseline, and sustains an aggregate throughput of $\approx 10^5$ queries per second on our setup. These results demonstrate that a layered routing strategy can deliver freshness, speed and cost-efficiency simultaneously in production-grade RAG systems.

## I. Introduction

Large-language models (LLM) have turned everyday chat interfaces into powerful knowledge tools, enabling products such as ChatGPT, virtual assistants, and automated customer-service agents [1]. However, enterprise deployment remains difficult because models trained on static snapshots cannot keep pace with continuously evolving document stores. Furthermore, naïve retrieval pipelines struggle to stay below sub-second latency under thousands of concurrent requests and GPU costs rise rapidly as knowledge bases scale to terabytes [2] [3]. Although Retrieval Augmented Generation (RAG) grounds an LLM in external context to address this bottleneck, existing RAG systems typically cannot optimize the three constraints all at once: time latency, scalability, and costs.

We therefore explore a middle ground with PentaRAG, a five-layer routing scheme that balances scalability, latency, and cost without adding new model weights. A lightweight decision router filters each query through progressively heavier paths: (1) a fixed key–value cache, (2) a semantic cache, (3) a memory-recall mode, (4) a session-level adaptive memory, and (5) a conventional retrieval-augmentation step, returning the first high-confidence answer it encounters. This design serves most requests from inexpensive cache or recall routes while still falling back to full retrieval when new information is required, reducing average latency and GPU usage compared with a single-stage RAG baseline on our enterprise-scale workload.

## II. Related Works

### 2.1 Parametric Memory in LLM

Early transformer models treat all knowledge as parametric memory information stored implicitly in their weights. Scaling this memory yields impressive few-shot competence: GPT-3 (175 B parameters) can answer questions and translate text without task-specific fine-tuning, largely by retrieving facts from its own weights [4]. However, because the training data are fixed at snapshot time, these models cannot incorporate new documents or correct outdated facts without full retraining, and context windows remain too small to "page in" long documents at inference.

Researchers therefore began to externalize memory. DeepMind's Retrieval-Enhanced Transformer (RETRO) attaches a frozen nearest-neighbor retriever to bring chunks from a 2-trillion-token datastore into the decoder, matching GPT-3 accuracy with 25 × fewer parameters [5]. More recently, MemGPT frames an LLM as an operating system that manages a hierarchy of fast and slow memory tiers, swapping content in and out of the context window to simulate an effectively unbounded working memory [6]. These efforts highlight the trade-off between parametric capacity and explicit memory access, motivating hybrid systems such as the five-layer cascade we propose in Section 3 that combine in-weight recall with multiple levels of external retrieval and caching.

**2.2 Retrieval Augmented Generation (RAG)**
The canonical RAG pipeline couples a neural retriever with a sequence-to-sequence generator so that the LLM can ground its output in external evidence. REALM first demonstrated this idea at pre-training time, back-propagating through a million-document index and showing large gains on open-domain QA with a 7 B-parameter model [7]. RAG generalized the recipe to downstream fine-tuning, proposing both token-wise and sequence-wise conditioning on retrieved passages and attaining state-of-the-art accuracy on NaturalQuestions and WebQuestions [8]. Follow-up work explored how to fuse multiple documents: Fusion-in-Decoder (FiD) re-encodes each passage separately and lets the decoder aggregate evidence, improving multi-hop reasoning tasks [9]. ATLAS pushed the approach to few-shot settings, jointly pre-training retriever and generator so that only 64 labelled examples are needed to surpass a 540 B-parameter baseline on NaturalQuestions [10]. Finally, hybrid designs such as HybridRAG/GraphRAG combine dense retrieval with knowledge-graph look-ups to boost factual precision in specialized domains like finance [11].

Despite these advances, most systems still rely on a single retrieval-generation path. They incur full retriever latency even for repeated or semantically similar queries and provide no mechanism for on-line knowledge growth. Section 3 introduces a five-layer cascade that retains the strengths of RAG while adding fast cache-based paths and an adaptive memory, addressing the latency and freshness gaps identified above.

**2.3 LLM Evaluation for RAG: Tonic Validate and RAGAS**
To assess a RAG-based LLM application, we adopt the open-source tool-chain Tonic Validate [12] and RAGAS [13]. Tonic Validate executes the full pipeline of retriever and answer generator on a labelled question-answering set and logs candidate passages as well as model outputs. RAGAS then computes sentence-level entailment and semantic-similarity scores, yielding the seven metrics defined as follows.

1) answer similarity: measures how closely the RAG system's response matches a reference answer on a 0–5 scale, typically as an integer; 2) augmentation precision: assesses how well the answer incorporates relevant retrieved context. Scored from 0 to 1, based on whether information from each relevant context appears in the response; 3) answer consistency: indicates the proportion of the answer supported by retrieved context. The LLM lists key points in the response and checks how many are traceable to the context (0–1 scale); 4) duplication metric: flags whether the answer contains repeated content. This is a binary metric: `1` if duplicates are present, `0` otherwise; 5) factual correctness: evaluates factual agreement between the generated response and reference using precision, recall, and F1. Scored from 0 to 1 based on matching claims. 6) faithfulness metric: measures how factually consistent the answer is with the retrieved context, scored from 0 to 1, with higher values indicating better alignment. It is defined as follows:

$$score = \frac{the\ count\ of\ statements\ backed\ by\ the\ context}{total\ number\ of\ LLM\ generated\ claims}$$

7) answer relevancy: scores how well the answer responds to the user input, rewarding relevance and completeness. It is defined as follows:

$$score = \frac{1}{N}\sum_{i=1}^{N}\cos(E_{g_i}, E_o)$$

Where:
- $E_{g_i}$: Embedding vector of the $i$-th generated question
- $E_o$: Vector representation of the user input
- $N$: Total generated questions

**2.4 Efficiency metrics: GPU time per query and Queries-per-second**
Enterprise services must quantify how much GPU time every answer costs and how many answers can be served each second. We therefore report two system-level metrics that appear in recent serving papers such as vLLM with PagedAttention [14] and that can be collected with simple `nvidia-smi` polling and locust step-load tests [15] [16].

GPU time per query is defined as the total GPU time that system spends to serve one query, expressed in seconds of single GPU. It is defined as follows:

$$GPU\ time\ /query = \sum_{each\ GPU}(wall\ time\ \times utilization)$$

Where $wall\ time \times GPU\ utilization\ percentage$ refers to the actual GPU seconds consumed.

Queries-per-second (QPS) or throughput is measured by running a step-load in locust in wall-clock time, which is defined as follows:

$$QPS = \frac{\#Query}{t_{elapsed}}$$

These two metrics are complementary. GPU-seconds-per-query captures per-answer cost even at low load, while QPS shows how the system scales under concurrency. Reporting both, as we do in the Section 3.4.3, allows fair comparison across caching and retrieval strategies.

## III. Implementation
**3.1 System Design**
PentaRAG is divided into three subsystems as depicted in Figure 1, which are instant access cache to retrieve quick answer, knowledge access retriever to get the context information, and LLM engine to generate the response. These three subsystems are orchestrated by the decision router to retrieve the fastest answer among the five modules that can be delivered to the user.

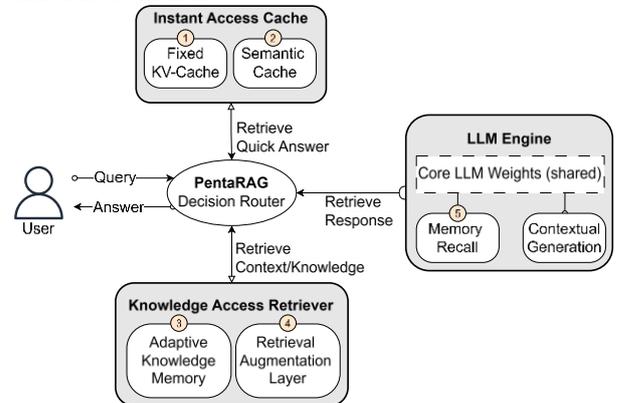

Figure 1 **Diagram flow of PentaRAG**. The **five numbered nodes (①–⑤)** constitute our **PentaRAG** methods.

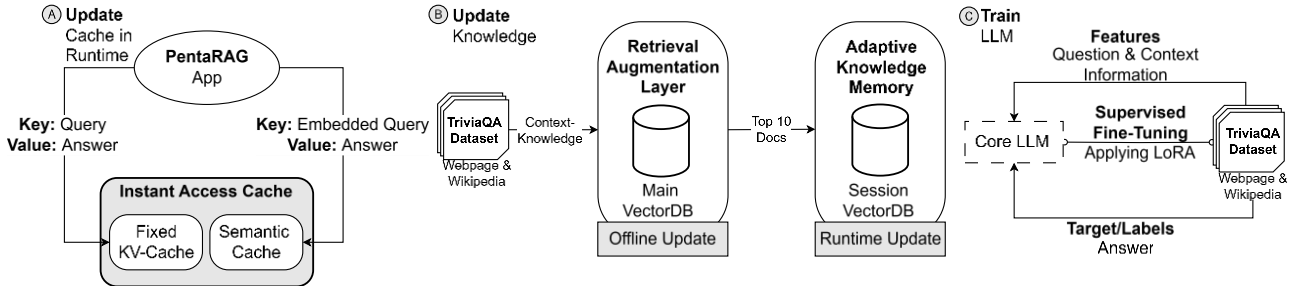

Figure 2 **Update and Training**. **(A)** Every served answer during runtime interaction is stored to the fixed KV-cache and semantic cache with the respective keys of string-formatted and vector embedding query. **(B)** Large repositories of context information are inserted offline in retrieval augmentation layer while the Adaptive Knowledge Memory will expand the knowledge incrementally by 10 top documents in runtime. **(C)** LLM is trained using supervised techniques to adapt with the domain knowledge.

Instant access cache refers to the cache layer that has a pair of question and answer history consisting of two modules: fixed KV-cache and semantic cache. During the runtime interaction with the user, if the passed query is 100% similar (the exact matched) with the previous query in the session history, the corresponding answer pair will be retrieved instantly with near-zero latency by the fixed KV-cache module. The same characteristic also applies to the semantic cache, but with the different query matching strategy. The semantic cache uses the cosine similarity to match the query with a certain threshold. If the two queries are matched semantically above the set threshold, the answer pair of the top-scored cosine distance will be retrieved.

Knowledge access retriever involves two modules: retrieval augmentation layer (naïve RAG) and adaptive knowledge memory. First, the query will be converted into the 1024-dimensional embedding vector space. Then, the query will be matched to the two modules that have different vector database collection using the cosine similarity. The retrieval augmentation layer serves as the main vector database retriever to access the large repositories of the set domain knowledge or context information containing billions of vectors. The adaptive knowledge memory, on the other hand, is a small subset of main vector database that is adaptively changing along the runtime session. We configure both vector database in a linear flat to fit the limited resource while maintaining a 100 % recall.

LLM Engine is the answer generator powered by the LLM that has two modules. Context generation module requires the context information to generate the response while the memory recall module does not. The Memory-Recall module lets the core LLM produce an answer directly from its internal knowledge; a confidence check then decides whether the answer is accepted or discarded. Because no external context is supplied, this mode relies entirely on the information the LLM absorbed during pre-training and domain fine-tuning.

### 3.2 Knowledge Maintenance Pipeline

Adaptation in PentaRAG proceeds on three complementary time scales. First, at request time (Fig. 2-A) each answered query is written back synchronously to both caches: the string key of the raw query populates the Fixed KV-Cache, while the same query's embedding key occupy the Semantic Cache. This zero-latency update guarantees that identical or semantically similar questions issued later will bypass retrieval entirely. Second, on a rolling basis (Fig. 2-B), the knowledge store behind the Retrieval Augmentation Layer is enlarged through two channels: (i) bulk offline ingestion of curated corpora into the main vector database and (ii) incremental online insertion of passages returned for every live query into an Adaptive Knowledge Memory shard. This dual feed keeps the retriever both comprehensive and freshly context aware. Third, in scheduled maintenance windows (Fig. 2-C) the accumulated ⟨question, context, answer⟩ triples are distilled into supervised training examples; a periodic fine-tuning step then refreshes the shared LLM weights so that newly introduced concepts migrate from retrieval into the model's own parametric memory.

### 3.3 Experimental Setup and Method
#### 3.3.1 Hardware Setup

All experiments were conducted on a single workstation equipped with an Intel Xeon Silver 4216 CPU (32 cores, 128 GB RAM) and five NVIDIA Quadro RTX 8000 GPUs (48 GB each). One GPU was reserved for supervised fine-tuning while the remaining four GPUs were allocated to the online service to host the fine-tuned LLM and the 1024-dimensional embedding model.

#### 3.3.2 Software Stack

| Stack | Function | Configuration |
|---|---|---|
| Milvus-DB [17] | Vector store for naïve RAG, adaptive knowledge memory, and semantic cache | flat index with cosine distance on 1024-D float32 vectors |
| VLLM [14] | Loads the LLM across multiple GPUs | tensor parallel size = 4; GPU memory utilization= 90% |
| Python dict cache | Fixed KV-Cache (exact string and answer pair) | In-memory, write-through on every response |
| Mxbai-embed-large-v1 | 1024-D embedding model for queries or passages [18] | HF hub id: mixedbread-ai/mxbai-embed-large-v1 |
| SFTT Trainer [19] | Supervised fine-tuning | • adam optimizer<br>• LR $2.5 \times 10^{-5}$<br>• linear scheduler<br>• 2 epochs |
| LoRA [20] | To reduce the number of trainable parameters | $r$=8, $\alpha$=16, dropout=0.05, causal-LM variant |
| Tonic Validate, RAGAS | Offline evaluation of fine-tuned checkpoints [12] [13] | Default RAGAS rubric; TriviaQA held-out split (100 samples) |

Table 1 **Software Setup**

All experiments ran on Ubuntu 22.04, CUDA 12.4, and cuBLAS 12.4 with the software setup as described in Table 1.

### 3.3.3 Offline vs Online Workflow

The evaluation pipeline is organized into three sequential phases, which are offline data ingest, supervised adaptation, and live serving. Each of which feeds the next to keep PentaRAG up-to-date while remaining fully reproducible.
1) **Offline ingest:** Bulk corpora (TriviaQA ± domain webpages) were embedded with mxbai-embed and loaded into the Main VectorDB.
2) **Supervised fine-tuning:** We fine-tuned the Ministral 8B (one of mistral model variants) on the training set of TriviaQA and selected the best checkpoint on a validation slice.
3) **Online service:** At runtime, queries first probe the Fixed KV-Cache and Semantic Cache (with a threshold of 85%); on a miss, the Retrieval Augmentation Layer (or Naïve RAG) fetches the top-3 passages, while the top-10 are asynchronously inserted into Adaptive Knowledge Memory.

### 3.3.4 Runtime Simulation

To evaluate PentaRAG performance during the runtime, we need to define a simulation to make the application not only giving the response to the user through Naïve RAG component but also utilizing other 4 components in a probabilistic manner. To realize this condition, we ran nine sessions of runtime with questions and contexts sampled from the Trivia-QA evaluation set. Each session started with an empty cache and consisted of a thousand questions drawn from either the dataset or the past questions. The likelihood of drawing a sample from past questions was controlled by the probability variable that was changing incrementally from 0 to 100% in each running question in the session. If the system logic decides to proceed with the past question, with 50% probability, the system will decide between two options: naively choosing the past question or slightly perturbing the past question.

### 3.4 Experimental Results
### 3.4.1 Supervised Fine-tuning

| Metric | Finetuned Mistral | Base Mistral | Improvement (%) |
|---|---|---|---|
| Answer Similarity | 4.070 | 3.770 | ▲7.958 |
| Augmentation Precision | 0.695 | 0.660 | ▲5.303 |
| Answer Consistency | 0.573 | 0.583 | ▼-1.749 |
| Duplication Metric | 0.300 | 0.060 | ▼-80.00 |
| Factual Correctness | 0.720 | 0.619 | ▲16.297 |
| Answer Relevancy | 0.895 | 0.823 | ▲8.738 |
| Faithfulness Metric | 0.884 | 0.874 | ▲1.129 |

Table 2 **Supervised Fine-tuning results on seven metrics**

Table 2 indicates that LoRA fine-tuning on TriviaQA improves the base mistral model checkpoint on five of seven metrics, with Answer Similarity up by 8 percent and Answer Correctness up by 16 percent. However, Answer Consistency falls slightly (about 2 percent) and the Duplication metric worsens, showing more repeated phrases. At the cost of a minor decline in stylistic diversity, the finetuned model is therefore markedly more accurate and relevant for the target domain that directly benefits the Memory Recall module, which draws its responses from the model's internal knowledge.

### 3.4.2 Simulated System Runtime

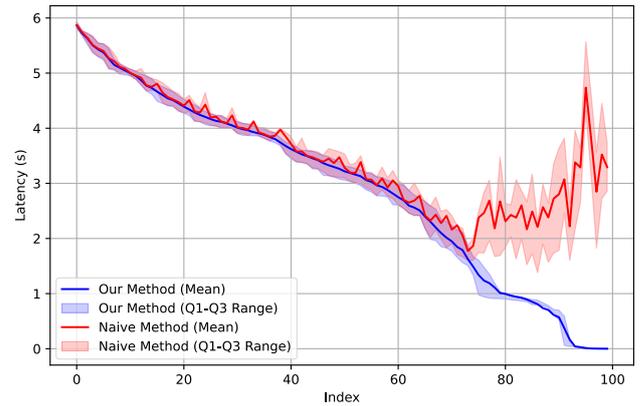

Figure 3 **Simulated runtime latency**. Mean latency (solid line) with Q1–Q3 band (shaded) averaged over nine 1,000-query sessions.

Figure 3 plots per-query latency as the system gradually "warms up." We replayed nine independent interaction sessions, each with 1,000 TriviaQA queries. From every session we retained the first 100 queries whose latency fell below six seconds, then sorted them in descending order of latency. At each index we averaged the nine latencies and shaded the inter-quartile range (Q1–Q3).

At the beginning the caches and Adaptive Knowledge Memory are empty, so the blue and red trajectories in Figure 3 almost coincide, although the naïve pipeline still shows slightly higher latency. As the session progresses the caches are populated with domain knowledge, latency decreases, and PentaRAG increasingly answers through the faster paths, such as cache and Memory Recall, rather than the naïve RAG retrieval pipeline.

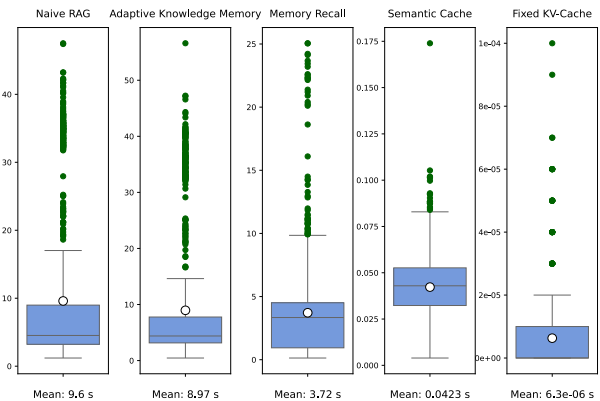

Figure 4 **Boxplot Latency Distribution of PentaRAG**. Boxes denote Q1–Q3, whiskers span the 5th–95th percentile, and outliers are plotted individually.

Our observation results (as depicted by Figure 4) show that the Naïve-RAG or Retrieval-Augmentation Layer exhibits the highest latency across all components. By contrast, both cache-based paths achieve almost zero latency, verifying the effectiveness of Fixed KV and Semantic caches for recurrent or semantically similar requests. The Memory-Recall mode is also moderately fast which can be considered as a practical fallback whenever the query is absent from cache but still within the model's parametric knowledge.

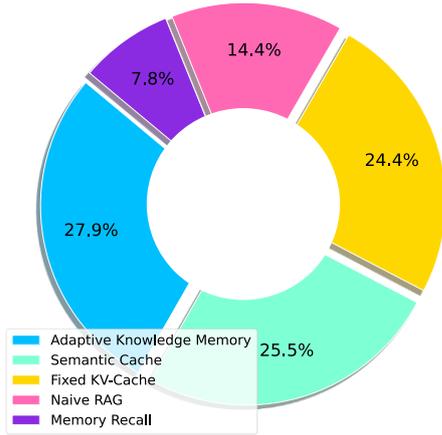

Figure 5 **PentaRAG Modules Usage Ratio**. Relative usage of each PentaRAG component during runtime, averaged over nine sessions (1,000 queries each).

To understand how often each path is actually utilized, Figure 5 plots the module-usage ratio accumulated over all nine evaluation sessions. Adaptive Knowledge Memory accounts for the largest share of traffic (27.9 %), followed by Semantic Cache (25.5 %) and Fixed KV-Cache (24.4 %). The heavyweight Naïve-RAG pathway and the Memory-Recall mode are invoked far less frequently (14.4 % and 7.8 %, respectively), confirming that once the caches warm, the system answers most queries through low-latency routes.

**3.4.3 Resource Efficiency Performance**

| Components | GPU Time/ query | Usage Ratio | Weighted GPU Time/query |
|---|---|---|---|
| Naive RAG | 0.53866 s | 14.4 % | 0.07757 s |
| Adaptive Knowledge memory | 0.53866 s | 27.9 % | 0.15029 s |
| Memory Recall | 0.25703 s | 7.8 % | 0.02005 s |
| Semantic Cache | 9.4e-04 s | 25.5 % | 2.4e-04 s |
| Fixed KV-Cache | 0 | 24.4 % | 0 |
| **Weighted Total** | - | 100 % | **0.24814 s** |

Table 3 **GPU cost per query for each PentaRAG component**

As shown in Table 3, retrieval-heavy routes dominate cost. Naïve-RAG and Adaptive Knowledge Memory each require around $0.54\ s$ per query which correspond to about 42% usage altogether of the total traffic. Semantic Cache needs just $9.4 \times 10^{-4}\ s$ and Fixed KV-Cache effectively does not cost the GPU time, while Memory-Recall sits between at $0.26\ s$. Weighting by the observed usage shares yields an overall mean of 0.248 GPU-$s$ per query, which is about twice more efficient than the Naïve-RAG baseline.

| Components | Query/sec (QPS) | Usage Ratio | Weighted QPS |
|---|---|---|---|
| Naive RAG | 0.45579 | 14.4 % | 0.06563 |
| Adaptive Knowledge memory | 1.81431 | 27.9 % | 0.50619 |
| Memory Recall | 2.51 | 7.8 % | 0.19578 |
| Semantic Cache | 208.33 | 25.5 % | 53.12415 |
| Fixed KV-Cache | 419,430 | 24.4 % | 102,341 |
| **Weighted Total** | - | 100 % | **≈102,395** |

Table 4 **Query-per-second throughput for each PentaRAG component**

Table 4 emphasizes each component's capacity as queries-per-second (QPS). Based on the measurement, cache-based paths dominate capacity: Fixed KV-Cache alone can serve more than $4 \times 10^5$ QPS and Semantic Cache around 200 QPS, together accounting for half the traffic. Memory-Recall delivers 2.5 QPS, whereas retrieval-based routes remain sub-QPS. Aggregating with the same usage weights gives an effective system throughput of $\approx 1 \times 10^5$ QPS, a three-order-of-magnitude gain over pure RAG.

These results confirm that once the caches warm, PentaRAG shifts most queries to near-free routes, halving GPU cost per answer and boosting overall throughput by more than thousand times relative to Naïve-RAG.

## V. Conclusion

This work introduced PentaRAG, a five-layer routing architecture that unifies key–value caching, semantic caching, in-weight memory recall, adaptive session memory, and traditional retrieval-augmented generation. By serving repeated and semantically similar queries from inexpensive cache or recall paths and reserving full retrieval only for novel questions, the system delivers sub-second latency at high concurrency while halving average GPU time per answer relative to a naïve RAG baseline. LoRA fine-tuning on TriviaQA further lifts semantic similarity and factual correctness without compromising retrieval faithfulness, demonstrating that parametric and non-parametric knowledge sources can be combined effectively. These results suggest that layered routing is a practical way to balance freshness, speed, and cost in enterprise LLM deployments. Future work will explore adaptive cache-retention policies and broader domain evaluations to confirm generality.

## Acknowledgement

This work was supported by the Institute for Information & Communications Technology Planning & Evaluation (IITP) grant funded by the Korea government (MSIT) under the project No. RS-2021-II211381 (Development of Causal AI through Video Understanding and Reinforcement Learning, and Its Applications to Real Environments) and No. RS-2022-II0951 (Development of Uncertainty-Aware Agents Learning by Asking Questions).

## REFERENCES

[1] OpenAI, "Introducing ChatGPT Enterprise," 28 August 2023. [Online]. Available: https://openai.com/index/introducing-chatgpt-enterprise/?utm_source=chatgpt.com. [Accessed 2 May 2025].

[2] promptingguide.ai, "Retrieval Augmented Generation (RAG) for LLMs," 25 April 2025. [Online]. Available: https://www.promptingguide.ai/research/rag. [Accessed 2 May 2025].

[3] J. Ligteringen, "Everything Wrong with Retrieval-Augmented Generation," [Online]. Available: https://www.leximancer.com/blog/everything-wrong-with-retrieval-augmented-generation. [Accessed 2 May 2025].

[4] T. e. a. Brown, "Language models are few-shot learners," *Advances in neural information processing systems,* vol. 33, pp. 1877-1901, 2020.

[5] S. e. a. Borgeaud, "Improving language models by retrieving from trillions of tokens," *International Conference on Machine Learning,* pp. 2206-2240, 2022.

[6] C. e. a. Packer, "MemGPT: Towards LLMs as Operating Systems," 2023.

[7] K. e. a. Guu, "Retrieval augmented language model pre-training," *International conference on machine learning,* pp. 3923-3938, 2020.

[8] P. e. a. Lewis, "Retrieval-augmented generation for knowledge-intensive nlp tasks," *Advances in neural information processing systems,* vol. 33, pp. 9459-9474, 2020.

[9] E. H. L. a. J. L. Choi, "Multi-Granularity Guided Fusion-in-Decoder," in *arXiv preprint arXiv:2404.02581*, 2024.

[10] G. e. a. Izacard, "Atlas: Few-shot learning with retrieval augmented language models," *Journal of Machine Learning Research,* vol. 24.251, pp. 1-43, 2023.

[11] B. e. a. Sarmah, "Hybridrag: Integrating knowledge graphs and vector retrieval augmented generation for efficient information extraction," *Proceedings of the 5th ACM International Conference on AI in Finance,* pp. 608-616, 2024.

[12] tonic.ai, "RAG metrics reference," 2024. [Online]. Available: https://docs.tonic.ai/validate/about-rag-metrics/tonic-validate-rag-metrics-reference. [Accessed 2 May 2025].

[13] RAGAS, 23 January 2025. [Online]. Available: https://docs.ragas.io/en/latest/concepts/metrics/available_metrics/#other-tasks. [Accessed 2 May 2025].

[14] W. e. a. Kwon, "Efficient memory management for large language model serving with pagedattention," in *Proceedings of the 29th Symposium on Operating Systems Principles*, 2023.

[15] NVIDIA, [Online]. Available: https://docs.nvidia.com/deploy/nvidia-smi/index.html. [Accessed 2 May 2025].

[16] Locust, "Running Locust in Step Load Mode," [Online]. Available: https://docs.locust.io/en/0.14.6/running-locust-in-step-load-mode.html. [Accessed 2 May 2025].

[17] e. a. Jianguo Wang, "Milvus: A Purpose-Built Vector Data Management System," *Proceedings of the 2021 International Conference on Management of Data,* pp. 2614-2627, 2021.

[18] e. a. Sean Lee, "Open Source Strikes Bread - New Fluffy Embeddings Model," 8 March 2024. [Online]. Available: https://www.mixedbread.ai/blog/mxbai-embed-large-v1. [Accessed 2 May 2025].

[19] e. a. Leandro von Werra, "TRL: Transformer Reinforcement Learning," Github, 2020. [Online]. Available: https://github.com/huggingface/trl. [Accessed 2 May 2025].

[20] E. J. e. a. Hu, "Lora: Low-rank adaptation of large language models," *International Conference on Learning Representations,* vol. 3, 2022.